\documentclass[12pt,draftclsnofoot,journal,onecolumn]{IEEEtran}

\usepackage{cite}
\usepackage{graphicx}
\usepackage{amsmath}
\usepackage{amsfonts}
\usepackage{array}
\usepackage{amssymb}
\usepackage{subfigure}
\usepackage[stable]{footmisc}
\usepackage{multirow}
\newtheorem{lemma}{Lemma}
\newtheorem{theorem}{Theorem}
\newtheorem{corollary}{Corollary}
\newtheorem{remark}{Remark}
\newtheorem{definition}{Definition}
\newtheorem{assumption}{Assumption}

\begin{document}

\title{Open vs. Closed Access Femtocells in the Uplink}
\author{Ping Xia, Vikram Chandrasekhar and Jeffrey G. Andrews
\thanks{P. Xia and J. G. Andrews are with the Wireless Networking and Communications
Group, Department of Electrical and Computer Engineering, The
University of Texas at Austin, 1 University Station C0803, Austin,
TX 78712. Email: pxia@mail.utexas.edu and jandrews@ece.utexas.edu. V. Chandrasekhar is now with Texas Instruments, Dallas, TX. Manuscript last modified: \today}
}
\maketitle

\begin{abstract}
Femtocells are assuming an increasingly important role in the
coverage and capacity of cellular networks. In contrast to existing
cellular systems, femtocells are end-user deployed and controlled,
randomly located, and rely on third party backhaul (e.g. DSL or cable
modem). Femtocells can be configured to be either \emph{open access} or
\emph{closed access}. Open access allows an arbitrary nearby cellular
user to use the femtocell, whereas closed access restricts the use of
the femtocell to users explicitly approved by the owner. Seemingly, the network operator would prefer an open
access deployment since this provides an inexpensive way to expand their
network capabilities, whereas the femtocell owner would prefer closed
access, in order to keep the femtocell's capacity and backhaul to
himself.  We show mathematically and through simulations that the reality is more complicated
for both parties, and that the best approach depends heavily on whether
the multiple access scheme is orthogonal (TDMA or OFDMA, per subband) or
non-orthogonal (CDMA). In a TDMA/OFDMA network, closed-access is
typically preferable at high user densities, whereas in
CDMA, open access can provide gains of more than 200\% for the \emph{home
user} by reducing the near-far problem experienced by the femtocell. The results of this paper suggest that the interests of the femtocell
owner and the network operator are more compatible than typically
believed, and that CDMA femtocells should be configured for open
access whereas OFDMA or TDMA femtocells should adapt to the cellular user density.
\end{abstract}


\section{Introduction}
Femtocell access points (FAPs), also known as Home NodeBs (HNBs), are short-range low-power and extremely low-cost base
stations with third party backhaul (e.g. DSL or cable
modem). They are usually deployed and controlled by end-users who desire better indoor signal transmission and reception.
With the help of such FAPs, the network operator is able to extend high quality coverage inside peoples' houses without
the need of additional expensive cellular towers. At the same time, FAPs offload traffic from the cellular network and
subsequently improve network capacity\cite{TN07, FF09, VC08CommMag}. Not surprisingly, two-tier femtocell networks --
that
is, a macrocell network overlaid with femtocell access points -- are under intense investigation and rapid
deployment\cite{DNK09CommMag, DNK09CommMag2, RYK09CommMag, PH09CommMag, MY09CommMag, DLP09CommMag}.

\subsection{Interference Management Issues in Femtocells}
Despite FAPs promise, many concerns still remain, especially cross-tier interference\cite{HC07PIMR, LH07PIMR, VC09TwcUP,
NSN073GPP}. Two particular aspects of FAPs give rise to serious interference issues: 1) the co-channel
spectrum sharing between femtocells and macrocells; 2) the ``random'' placement of FAPs. First, unlike Wi-Fi access
points, FAPs serve users in licensed spectrum, to guarantee Quality-of-Service (QoS) and because the devices they
communicate with are developed for these frequencies. Compared to allocating separate channels inside
the licensed spectrum exclusively to FAPs, sharing spectrum would be preferred from an operator perspective
\cite{VC08CommMag,DC08Glob}. Secondly, FAPs are installed by end-users in a ``plug-and-play'' manner, which
translates into ``randomness'' in their locations: they can be deployed anywhere inside the macrocell area with no
prior warning\cite{VC08CommMag,Aricent08}. For these two reasons, interference in two-tier networks is quite different
than in conventional cellular networks, and endangers their successful co-existence\cite{VC09TwcUP,JS97IPC,DLP10EURASIP}. A
typical scenario is the ``Dead Zone'' or ``Loud Neighbor'' problem, where mobile users transmit and receive
signals at positions near FAPs but far from the macrocell BSs, causing significant macro-to-femto interference in the
uplink. In the downlink, these users likewise suffer from low signal to interference ratios (SIRs) because of the strong interference from the FAPs.  These affects are akin to the well known near-far problem, but exacerbated by the de-centralization and lack of coordinated power control inherent in a two-tier network.

Because of the presently non-existent coordination between FAPs and macrocell BSs, centralized cooperation to mitigate cross-tier interference is infeasible in the near future, and so in this paper we assume that a two-tier network needs to adopt decentralized strategies for interference management\cite{VC09TwcSP, VC09TwcCIT, VC09TwcPC} such as femtocell access control\cite{VC09TwcUP, DC08Glob, GDLR10CommMag,DLP10EURASIP}. Femtocell access control schemes can be divided into two categories: closed and open access. FAPs only provide service to specified subscribers in closed access, to ensure they can monopolize their own femtocell and its backhaul with privacy and security. However this potentially leads to severe cross-tier interference as described above. On the contrary, open access allows arbitrary nearby cellular users to use the femtocell. Seemingly, open access is beneficial to network operators, by providing an inexpensive way to expand their network capacities by leveraging third-party backhaul for free. Open access also reduces macro-to-femto interference by letting strong interferers simply use the femtocell and coordinate with the existing users through it. However, in order to attain a certain target receive power at the FAP, the handed off cellular user in open access generally transmits
with higher power to the FAP (thereby creating increased femto-to-macro interference) as compared to the in home user.
Thus open access potentially deteriorates QoS provided to cellular users remaining in the macrocell (arising due to
increased interference). Crucial unanswered questions remain in femtocell access control, such as:
\begin{enumerate}
\item Which mode meets the interests of femtocell owners? Which mode is preferable to the network operator? Are these
    two choices the same or different?
\item How does the answer depend on factors such as multiple access protocol (e.g. OFDMA, CDMA), user densities, user scheduling, and femtocell backhaul constraints?
\end{enumerate}

\subsection{Related Work}
The uplink interference in two-tier femtocell networks was evaluated in \cite{VC09TwcUP}, showing that tier-based open access can reduce the interference and offer an improvement in the network-wide area spectral efficiency -- the feasible number of femtocells and macrocell users per cell-site. Similar conclusions were presented in
many simulation-centric studies accomplished by the 3GPP RAN 4 group\cite{NSN073GPP,NV073GPP,3GPP}. Downlink network capacities under open and closed access were explored in \cite{NV073GPP}; Feasible combinations of femtocells and
macrocells under the constraint of network interference were examined in \cite{NSN073GPP}; Various scenarios were presented in \cite{3GPP} to compare femtocell open and closed access. All these simulations show that with adaptive open access, the interference in two-tier networks is mitigated and the deployment of co-channel femtocells
becomes feasible. However, since femtocells are installed and paid for by their owners, it is necessary to evaluate their loss of femtocell resources in open access. It is important that the benefits of mitigated interference are not undermined by the loss of femtocell resources, such as over-the-air (OTA) and backhaul capacity.

The issues of femtocell backhaul sharing in open access were examined in \cite{DC08Glob}, which simulated open and closed
access in HSDPA, with the thesis that completely open access is problematic because of sharing limited femtocell
backhaul among a potentially large number of mobile users. Based on simulations incorporating femtocell backhaul issues
and cross-tier interference, this work concludes that open access with a restriction on the number of supported users at
the FAP is the preferred approach. We derive the same conclusion in uplink based on both analytical and simulations
results. Moreover, we show that such conclusion strongly depends on whether the multiple access scheme is orthogonal (TDMA or OFDMA) or non-orthogonal (CDMA).

The increased handover frequency and hence overhead signaling in open access is a possible challenge to its implementation. A technique combining intracell handovers with power control was proposed in \cite{DLP10EURASIP}, and a \emph{hybrid access} model -- open access with a cap on the amount of resources allocated to the cellular users -- was simulated in \cite{GDLR10CommMag}. Both of these approaches substantially reduce the number of handovers in open access while mitigating the cross-tier interference. In this paper, we simply call the hybrid access model open access, since our open access approach has an upper limit of $K$ users, where $K$ could become arbitrarily large to conform to fully open access.

\subsection{Contributions}
This paper evaluates the performance of femtocell open and closed access in the uplink, from the viewpoints of both the
femtocell owner (owner's achieved rate) and the network operator (cellular users' sum throughput).

First, we derive the cumulative distribution function (CDF) for uplink cross-tier interference in
orthogonal multiple access schemes (TDMA or OFDMA). The capacity tradeoffs for both the femtocell owner and cellular users are then presented. In TDMA or OFDMA, the preferences of the femtocell owner and the network operator are highly dependent on cellular user density: Their choices are incompatible/ open access/ closed access in low/ medium/ high
cellular user density respectively. Thus, for 4G networks (LTE \& WiMAX) that use OFDMA, the femtocell access control should likely be adaptive to the cellular user density.

Second, by deriving lower bounds on the performance of open access, we show that in non-orthogonal multiple access (i.e.
CDMA) open access is a strictly better choice for the home user. In typical propagation scenarios, it provides more than a factor of $3$ rate gain to the femtocell owner by lowering interference. From the viewpoint of the network operator, open access is also preferred. In the less important regime of low cellular user density, open access achieves almost the same performance as closed access, while in the important regime of high cellular user density, it improves performance significantly. We suggest that femtocell open access is the preferred approach for CDMA  femtocells (i.e. 3G), from the viewpoints of both femtocell owners and network operators.

The rest of this paper is organized as follows. In Section II, we describe the system model and assumptions in detail. The capacity contours in orthogonal multiple access schemes are presented in Section III, which in non-orthogonal multiple access scheme are derived in Section IV. Numerical results are used to illustrate the main takeaways in Section V.

\section{System Model}
In the interior area of a macrocell of radius $R$, the macrocell BS is located at the center, with a single femtocell
access point (FAP) at a distance $D$ away from it. In multiple femtocells scenarios, the uplink femto-to-femto interference does not change much between closed and open access, as long as the density of femtocells is not very high. Thus it is reasonable to consider a single femtocell, to focus our scope on the mitigation of cross-tier interference in open vs. closed access, which indeed is the main motivation of access control mechanisms. Suppose there are $N$ cellular users, denoted as ${U_{1},
U_{2},\ldots, U_{N}}$, roaming inside the macrocell. Their positions are i.i.d. random variables, uniformly distributed in the macrocell area. The femtocell owner, or alternatively the home user, is
denoted as $U_{0}$. Since the home user
is transmitting and receiving inside the small area of a house, we could assume it is located at a \emph{deterministic}
position, with a distance of $d$ from the FAP. As the subscriber, the home user would always talk to the FAP. On the other hand, cellular users can be served by the macrocell BS, or the FAP if in femtocell open access.

\subsection{Channel Model and Interference}
We consider path loss attenuation effects only and ignore short-term fading in our channel model. This assumption is
reasonable because fading does not have a large effect in a wideband system with sufficient diversity, e.g. RAKE
receiver (CDMA), or multi-antenna diversity or distributed subcarrier allocation (OFDMA). The path-loss exponent of
outdoor (indoor) transmission is denoted by $\alpha$ ($\beta$). In particular, the channel model is given by
\begin{equation}
H(|x|)=
\begin{cases}
|x|^{-\alpha} & \mbox{outdoor \& cross-wall transmission}\\
|x|^{-\beta} & \mbox{indoor transmisstion}
\end{cases}
\end{equation}
Here, $|x|$ is the distance from the transmitter to the respective base station. Setting $\alpha > \beta$ incorporates wall penetration loss in our channel model.

\begin{assumption} \label{majorAS}
\emph{We assume that there is no coordination between the FAP and the macrocell BS, nor between different FAPs, in terms of power control or resource scheduling.}
\end{assumption}

In the uplink, denote $P_c$ and $P_f$ as the received power at the macrocell BS and the FAP respectively. Through
uplink power control, a macrocell user $U_j$ causes interference of $P_c h_j/g_j$ to the FAP, where $h_j$ ($g_j$) is its channel to the FAP (the macrocell BS). Conversely, a femtocell user $U_i$ causes interference of $P_f g_i/h_i$ to the macrocell BS.
\begin{definition} \label{DefI}
For $U_{j}$ $\in$ $\{U_{0}, U_{1}, \ldots, U_{N}\}$, its \emph{interference factor} $I_j$ is defined as $h_j/g_j$. $\{I_{1},I_{2},\ldots, I_{N}\}$ are i.i.d. random variables, and we define their ordered statistics as
\begin{align}
I_{(1)}=\min(I_1, \ldots, I_N), \ldots, I_{(k)}= \min(\{I_1, \ldots, I_N\}\backslash\{I_{(1)}, \ldots, I_{(k-1)}\}),
\ldots, I_{(N)}=\max(I_1, I_2, \ldots, I_N) \nonumber
\end{align}
where ``$\backslash$'' means the set subtraction operation. Corresponding to the ordered statistics of \emph{interference factors}, cellular users are reordered as $\{U_{(1)}, U_{(2)}, \ldots,U_{(N)}\}$.
\end{definition}
\begin{assumption} \label{AS1}
\emph{We assume $I_0 \geq I_{(N)}$ holds, because the home user is closer to the FAP, and the indoor channel has a smaller path loss exponent.}
\end{assumption}
\begin{lemma}\label{cdf}
\emph{The cumulative distribution function for the interference factor of each cellular user is
\begin{align}
F_I(i)&=
\begin{cases}
(r/R)^2 & 0 \leq i <(\frac{R}{R+D})^{\alpha}\\
L(i) &(\frac{R}{R+D})^{\alpha}
\leq i < 1 \\
\frac{\pi - \varphi + 0.5 \sin(2\varphi)}{\pi} & i=1\\
1-L(i) & 1 < i \leq
(\frac{R}{R-D})^{\alpha} \\
1-(r/R)^2 & (\frac{R}{R-D})^{\alpha} < i
\end{cases}
\end{align}
where r, L(i), $\varphi$ are given by:
\begin{align}
r&=\frac{i^{1/\alpha}D}{|1-i^{2/\alpha}|}, \quad \varphi
=\arccos\left(\frac{D}{2 R}\right), \nonumber \\
L(i)&=\frac{\pi(R^2+r^2)-(\pi-\theta+0.5 \sin2\theta)r^2-(\pi -\phi+0.5 \sin2\phi)R^2}{\pi R^2}, \nonumber
\end{align}
where $x_c=\frac{ D i^{2/\alpha}}{|1-i^{2/\alpha}|}$ , $\theta=\arccos\left(\frac{r^2+x_c^2-R^2}{2 r x_c}\right)$ and
$\phi=\arccos\left(\frac{x_c^2+R^2-r^2}{2 x_c R}\right)$.
}
\end{lemma}
\begin{IEEEproof}
See Appendix \ref{cdfproof}
\end{IEEEproof}

In non-orthogonal multiple schemes (CDMA), the interference is additive. For a set of $k$ \emph{interference factors} $\{I_{n_1}, I_{n_2}, \ldots, I_{n_k}\}$, denote function $G_I(k,\cdot)$ as the CDF of $\sum_{j=n_1}^{n_k} I_j$, which is given by
\begin{displaymath}
G_I(k,i)=\mathbb{P} (\sum_{j=n_1}^{n_k} I_j \leq i).
\end{displaymath}
$G_I(k,\cdot)$ is the same for whatever $k$ \emph{interference factors}, since the \emph{interference factors} are
i.i.d..
\begin{lemma}
\emph{An upper bound on the CDF function $G_I(k,\cdot)$ is given by
\begin{align}
G_I (k,i) \leq G_I^{ub}(k,i)=\left(F_I(i)\right)^k \triangleq F_I^k(i)
\end{align}
}
\end{lemma}
\begin{IEEEproof}
Consider a ``dominant interferer'', i.e. an interferer who alone causes interference sufficient for outage. Without loss of generality, we assume $I_{n_1}$ is the maximum \emph{interference factor} of
$\{I_{n_1},\ldots, I_{n_k}\}$. Therefore $\mathbb{P} (I_{n_1} \leq i) = F_I^k(i)$, which provides an upper bound on $G_I(k,\cdot)$.
\begin{align}
G_I(k,i)&= \mathbb{P} \left(\sum_{j=n_1}^{n_k} I_j \leq i \right) = 1- \mathbb{P}\left(\sum_{j=n_1}^{n_k} I_j > i \right)
\leq 1-\mathbb{P}\left(I_{n_1} >i \right) = F_I^k(i)
\end{align}
\end{IEEEproof}

To summarize, in this subsection we derived the CDF $F_I(\cdot)$ for each \emph{interference factor}, which will be used
to calculate outage probability for orthogonal multiple access. The CDF $G_I(\cdot,\cdot)$ for the sum of these
\emph{interference factors} was also derived, and will be used to calculate outage probability for non-orthogonal multiple access.

\subsection{Hand off Metric and Procedure}
When the FAP deploys open access control, it can choose to serve cellular users based on certain metrics. A typical metric is that
it provides service to cellular users if both of the following two conditions hold: 1) these cellular users cause
outage to the home user, and 2) the FAP has available resources. Such a metric allows cellular users to share
the femtocell resources when they can potentially boost the capacity of femtocell owners by reducing co-channel
interference. Suppose the maximum number of additional cellular users that the FAP can serve is $K$.
\begin{assumption} \label{AS2}
\emph{When cellular users cause outage to the home user, the FAP picks the most noisy interferer from
the macrocell to serve. This hand off procedure continues as long as the home user still experiences outage and the number of handed off cellular users does not exceed $K$. }
\end{assumption}
Based on this assumption, when the FAP provides service to $L$ cellular users, these users must be the strongest
interferers $\{U_{(N)}, U_{(N-1)}, \ldots, U_{(N-L+1)}\}$, which reduce the macro-to-femto interference to the largest
extent possible. When served by the FAP, these cellular users cause interference of $\{P_f/I_{(N)}, P_f/I_{(N-1)},
\ldots, P_f/I_{(N-L+1)}\}$ to the macrocell respectively, which are also the smallest possible. This assumption is of
significant importance from both practical and analytical aspects. In practice, it meets the interests of both the femtocell owner and the network operator by maximally reducing the interference. Additionally, it ensures that the open access admission control in our paper is optimal.

\subsection{Resource Allocation and Ergodic Rate} \label{subs1}
From Assumption \ref{majorAS}, we consider distributed resource allocation in two-tier femtocell network as follows.

\textbf{Backhaul Allocation.} The macrocell BS usually has a large backhaul capacity. So when a cellular user is served by the macrocell BS, its rate will not be constrained by the backhaul. However, the FAP backhaul capacity, denoted as $\mathcal{C}_b$, is typically modest and often shared, common examples being DSL and cable modem. Thus it is necessary to incorporate the FAP's backhaul allocation into the analysis. As the FAP serves additional $L$ ($0 \leq L \leq K$) cellular users, the home user is allocated with a portion $\lambda_L$ of this backhaul capacity, while each of the $L$ cellular users is assigned a portion $\mu_L$ of the FAP backhaul capacity. Obviously, we have
\begin{align}
\lambda_L + L \mu_L \leq 1, \quad \lambda_L, \mu_L \geq 0
\end{align}
In both closed and open access, when the FAP does not serve any cellular users, there is no backhaul issue, and $\lambda_0=1$ and $\mu_0=0$.
\begin{assumption}
\emph{For} $\{\lambda_0, \lambda_1, \ldots, \lambda_K\}$ \emph{defined above, the following inequality holds:}
\begin{align}
\lambda_0 \geq \lambda_1 \geq \ldots \geq \lambda_K
\end{align}
\emph{Because as users are added, the fraction of resources allocated to the home user should not increase.}
\end{assumption}

\textbf{Time Allocation.} A two-tier network needs to deal with the issue of time
allocation in orthogonal multiple access schemes (TDMA or OFDMA per subband).
In the macrocell network, since all the users
have the same rate requirement and they are i.i.d. located
inside the macrocell, the time resources will be fairly
allocated among them. Therefore, when the macrocell BS is serving
$M$ cellular users ($M \leq N$, since some cellular users can
talk to the FAP), each user enjoys an average time fraction\footnote{Although in practice the
time slot assigned to a mobile user can only
be a group of discrete values, the average time fraction of
the mobile user can be any value between $0$ and $1$. The same
argument is applicable to the value of $\lambda_L$ and $\mu_L$.} of $1/M$. In the femtocell network, when the FAP serves additional $L$ cellular users, the time fraction allocated to the home user and each of the $L$ cellular users should be $\lambda_L$ and $\mu_L$ respectively, according to the allocation in backhaul capacity.

\begin{assumption}
\emph{Each macrocell user has a rate requirement $\mathcal{C}$, while each femtocell user has a rate requirement of $\min\{\mathcal{C}, \lambda \mathcal{C}_b\}$, where $\lambda \mathcal{C}_b$ is its allocated backhaul capacity. Correspondingly, each user has a SIR target. The user achieves its required rate when the received SIR at or above its SIR target\footnote{As cellular networks are strongly interference limited, thermal noise is negligible.}. Otherwise it is in outage and the rate is zero.}
\end{assumption}
\begin{definition} \label{DEF3}
The event $A_{L} (L \in \{0,1,2,\ldots,K\})$ is defined as the FAP provides service to $L$ additional cellular users. In event
$A_L$,
denote the SIR targets of the home user, handed off $L$ cellular users and the remaining $N-L$ cellular users as
$\Gamma_{f,L}$, $\Gamma_{h,L}$ and $\Gamma_{c,L}$ respectively. Their success probabilities are denoted as $p_{f,L}$,
$p_{h,L}$ and $p_{c,L}$ accordingly.
\end{definition}

\begin{definition} \label{DEF4}
The \emph{ergodic rate} for a mobile user $U_j$ is its rate requirement multiplied by its success probability.
\end{definition}

In this paper, we evaluate open vs. closed access from the viewpoints of the femtocell owner -- the home user's \emph{ergodic rate} $\mathcal{C}_0$, and the network operator -- cellular users' \emph{sum throughput} $\mathcal{C}_{sum}$, which is defined as the sum of all cellular users' \emph{ergodic rates}. Although we use the hybrid model in \cite{GDLR10CommMag} as a more general form of open access, the overhead signaling from handovers still would affect the rates of mobile and femtocell users. However, since it is difficult to quantify precisely, often involves separate overhead channels, and the exact implementation varies significantly from protocol to protocol, we do not include the impact of handover signalling in the analysis.

\section{Capacity Contours in Orthogonal Multiple Access Schemes}
In this section, we analyze a TDMA scenario, which can also be viewed as OFDMA on a per subband basis, since each
subband is orthogonal and allocated in a TDMA fashion. We first consider the scenario when the FAP can serve $K$
cellular users. We then focus on the important special case of $K=1$.

In an arbitrary time slot, suppose users $U_i$ and $U_j$ are active at the femtocell and the macrocell respectively. Therefore the received SIRs at the two base stations are
    \begin{align}
    \begin{cases}
    \frac{P_f}{P_c I_j}  & \mbox{SIR at the FAP} \\
    \frac{P_c}{P_f/I_i}  & \mbox{SIR at the macrocell BS}
    \end{cases}
    \end{align}
\begin{theorem} \label{TCA}
\emph{In TDMA or OFDMA, the home user's ergodic rate and the cellular users' sum throughput in femtocell closed access
are
given by}
\begin{align}
\mathcal{C}_0= \min(\mathcal{C}, \lambda_0 \mathcal{C}_b) F_I\left(\frac{P_f}{P_c \Gamma_{f,0}}\right), \quad
\mathcal{C}_{sum}=N \mathcal{C} \mathbb{P}\left(\frac{P_c}{P_f/I_0} \geq \Gamma_{c,0}\right)
\end{align}
\end{theorem}
\begin{IEEEproof}
Since at the macrocell BS, each cellular user causes interference to the home user during its time slot, namely $1/N$,
the \emph{ergodic rate} of the home user is
\begin{flalign}
\mathcal{C}_0 =\min(\mathcal{C}, \lambda_0 \mathcal{C}_b) \sum_{j=1}^N \frac{1}{N} \mathbb{P}\left(\frac{P_f}{P_c I_j}
\geq \Gamma_{f,0}\right)
=\min(\mathcal{C}, \lambda_0 \mathcal{C}_b) F_I\left(\frac{P_f}{P_c \Gamma_{f,0}}\right)
\end{flalign}
On the other hand, each cellular user experiences an interference of $P_f/I_0$ from the home user. Their sum rate is
\begin{align}
\mathcal{C}_{sum}&=N \mathcal{C} p_{c,0}= N \mathcal{C} \mathbb{P} \left( \frac{P_c}{P_f/I_0} \geq \Gamma_{c,0}\right)
\end{align}
\end{IEEEproof}

\begin{remark} \label{reTCA}
Theorem \ref{TCA} shows that the home user's rate is independent of the number of cellular users -- the goal of closed
access. In TDMA the interference is time shared, so the average outage probability of home
user is not scaled by $N$. Things are different in CDMA, as shown in the next section.
\end{remark}

It is important that the SIR target of cellular users in the macrocell (in both open and closed access) is
an increasing function of their density. Intuitively, when the macrocell BS serves more mobile users, each of them has a
smaller time fraction and must increase its SIR target to achieve a given rate requirement.

In closed access, since the received SIR of a cellular user in the macrocell is a constant value of $\frac{P_c}{P_f/I_0}$, there is a cutoff user loading $N_c^*$, such that: 1) when $N \leq N_c^*$, each cellular user's SIR target
constraint is satisfied and their
\emph{sum throughput} is the maximum possible, $\mathcal{C}_{sum}=N \mathcal{C}$; 2) when $N > N_c^*$, each cellular
user's SIR target is infeasible and  $\mathcal{C}_{sum}=0$. The value of $N_c^*$ is governed by the inequality $\frac{P_c}{P_f/I_0} \geq \Gamma \mbox{ (SIR target)}$, which for example in a Gaussian channel is $N_c^*=\lfloor\frac{1}{\mathcal{C}} \log_2(1+\frac{P_c}{P_f/I_0})\rfloor$.

It is less clear if there is such a cutoff value $N_o^*$ in open access, because the received SIR of each cellular user in the macrocell is not constant, due to the random interference from the cellular users supported at the FAP. The
simulation results in Section V show that such a cutoff value $N_o^*$ occurs under practical network configurations, which
is essentially due to two particular aspects of open access. First, the FAP predictably allocates a large portion of OTA
resources to the home user. Therefore the femto-to-macro interference is still a constant value for a large portion of
time. Second, the cellular users served by the femtocell must be very close to the FAP according to the handoff criteria, which greatly reduces the randomness of their location. As a result, their interference to the macrocell is also nearly deterministic. The numerical relation between $N_c^*$ and $N_o^*$ is discussed later in Remark \ref{recutoff}.

In femtocell open access, events $\{A_L, L=0,1,2,\ldots,K\}$ can occur. From Assumption \ref{AS2}, FAP will
pick the strongest interferers sequentially, so
\begin{align}
\mathbb{P} (A_L) =
\begin{cases}
\mathbb{P} \left(\bigcap\limits_{j=0}^{L-1}\left\{\frac{P_f}{P_c I_{(N-j)}} < \Gamma_{f,j}\right\}, \frac{P_f}{P_c
I_{(N-L)}} \geq \Gamma_{f,L}\right) & L<K \\
\mathbb{P} \left(\bigcap\limits_{j=0}^{K-1} \left\{\frac{P_f}{P_c I_{(N-j)}} < \Gamma_{f,j}\right\}\right) & L=K\\
\end{cases}
\end{align}
\begin{lemma} \label{TOA}
\emph{In TDMA or OFDMA, the ergodic rate of the home user and sum throughput of the cellular users in open access are
given by}
\begin{flalign} \label{equlemma3}
\mathcal{C}_0=\sum_{L=0}^K \min(\mathcal{C}, \lambda_L \mathcal{C}_b) p_{f,L} \quad \quad
\mathcal{C}_{sum}&=N \mathcal{C} p_{c,0}+ \sum_{L=1}^K \left\{(N-L) \mathcal{C} p_{c,L} + L \min(\mathcal{C}, \mu_L
\mathcal{C}_b) p_{h,L}\right\}
\end{flalign}
\emph{where $p_{f,L}$,$p_{h,L}$ and $p_{c,L}$ are success probabilities for the home user, the supported cellular
users
at the femtocell and the remaining cellular users at the macrocell respectively, which are given by}
\begin{align} \label{equTPi}
p_{f,L}&=
\begin{cases}
\mathbb{P} (A_L) & L=0,1,\ldots,K-1 \\
\frac{1}{N-K} \sum_{j=1}^{N-K}\mathbb{P} \left(\frac{P_f}{P_c I_{(j)}} \geq \Gamma_{f,K}, A_K \right) & L=K\\
\end{cases}\\
p_{h,L}&=\frac{1}{N-L} \sum_{j=1}^{N-L} \mathbb{P} \left(\frac{P_f}{P_c I_{(j)}} \geq \Gamma_{h,L} , A_L \right)\\
p_{c,L}&=\lambda_L \mathbb{P} \left(\frac{P_c}{P_f/I_0} \geq \Gamma_{c,L} \right) \mathbb{P} (A_L) + \mu_L
\sum_{j=N-L+1}^{N} \mathbb{P} \left(\frac{P_c}{P_f/I_{(j)}} \geq \Gamma_{c,L} , A_L \right)
\end{align}
\end{lemma}
\begin{IEEEproof}
The key step in the proof is the calculation on success probabilities. Denote $S_f$ as the event that the home user succeeds in its communication process. When $L < K$, we have
\begin{align}
p_{f,L}=\mathbb{P} (S_f, A_L)=\mathbb{P} (S_f |A_L) \mathbb{P} (A_L)=\mathbb{P} (A_L)
\end{align}
When the FAP serves only $L$ ($L < K$) cellular users, it implies the home user
experiences no outage at this point based on Assumption \ref{AS2}. Therefore $\mathbb{P} (S_f |A_L)=1$ and the last
equality holds. When $L=K$, the remaining $N-K$ cellular users in the macrocell, which are $\{I_{(1)}, I_{(2)},\ldots,
I_{(N-K)}\}$, can possibly cause outage to the home user. Since they are fairly scheduled, they are equally
likely interfering the home user with probability $\frac{1}{N-K}$. So in the event of $A_K$, the probability of success
of the home user is
\begin{align}
p_{f,K}= \sum_{j=1}^{N-K} \frac{1}{N-K} \mathbb{P}\left(\frac{P_f}{P_c I_{(j)}} \geq \Gamma_{f,K}, A_K \right)
\end{align}
Similar arguments hold for $p_{h,L}$ and $p_{c,L}$.
\end{IEEEproof}

In open access, due to the random macro-to-femto and femto-to-macro interference, the cellular users' \emph{sum
throughput} is strictly between $0$ and $N\mathcal{C}$, which are two possible \emph{sum throughput} in closed access. Therefore, the network operator's choice between open vs. closed access is fairly clear.
\begin{remark} \label{resumT}
According to the value of $N_c^*$, the network operator prefers closed access when $N \leq
N_c^*$, while embracing open access when $N > N_c^*$.
\end{remark}

The reason why open access reduces the \emph{sum throughput} when $N \leq
N_c^*$ is explained as follows. The femto-to-macro interference is $P_f/I_0$ in closed
access for all time slots, which in open access after handoff will increase to (due to Assumption \ref{AS1}) $P_f/I_{(i)}$
in the time slot of $U_{(i)}$, the cellular user served by the FAP. The increased femto-to-macro interference from the
handed off cellular users indeed bottlenecks the performance of open access by reducing \emph{sum throughput}.
\begin{remark} \label{recutoff}
When the amount of cellular users in the macrocell is over $N_c^*$, the femto-to-macro interference in closed access
causes their \emph{sum throughput} to be zero, which should also be true in open access due to the increased
femto-to-macro interference. Considering the at most $K$ cellular users served by the FAP, the cutoff value $N_o^*$
should be given by $N_o^* \leq N_c^*+K $.
\end{remark}

In the following, we focus on a special case of $K=1$. Such a case is important because femtocell owners can be reasonably expected as selfish users with their infrastructure.
\begin{theorem}\label{TOA1}
\emph{In TDMA or OFDMA, when the FAP is set to serve at most one cellular user, namely $K=1$, the
ergodic rate of home user and the sum throughput of cellular users in open access are given by}
\begin{align}
\mathcal{C}_0 &=\min(\mathcal{C}, \lambda_0 \mathcal{C}_b)F_I^N(\frac{P_f}{P_c \Gamma_{f,0}})+
\min(\mathcal{C},\lambda_1 \mathcal{C}_b) \frac{N}{N-1} F_I(\frac{P_f}{P_c \Gamma_{f,1}}) \left( 1-
F_I^{N-1}(\frac{P_f}{P_c \Gamma_{f,0}})\right) \nonumber \\
\mathcal{C}_{sum} &= N \mathcal{C} \mathbb{P}\left( \frac{P_c}{P_f/I_0} \geq \Gamma_{c,0} \right) F_I^N(\frac{P_f}{P_c
\Gamma_{f,0}}) + \mathcal{C}(N-1)p_{c,1}+ \min(\mathcal{C},\mu_{1}\mathcal{C}_b)p_{h,1}
\end{align}
\emph{Where $p_{c,1}$ and $p_{h,1}$ are given by}
\begin{align}
p_{c,1} &=\lambda_1 \mathbb{P} (\frac{P_c}{P_f/I_0} \geq \Gamma_{c,1}) \left(1-F_I^N(\frac{P_f}{P_c \Gamma_{f,0}})\right)
+ \mu_1 \min\left(1-F_I^N(\frac{P_f}{P_c \Gamma_{f,0}}), 1-F_I^N(\frac{P_f \Gamma_{c,1}}{P_c})\right) \nonumber \\
p_{h,1}&=\frac{N}{N-1} F_I(\frac{P_f}{P_c \Gamma_{h,1}}) \left( 1- F_I^{N-1}(\frac{P_f}{P_c \Gamma_{f,0}})\right)
\end{align}
\end{theorem}
\begin{IEEEproof}
See Appendix \ref{TOA1proof}
\end{IEEEproof}

Note that the SIR target of a femtocell user is a non-increasing function of its allocated time fraction. For example, in a Gaussian Channel, a femtocell user $U_{i}$ has a SIR target $\Gamma=2^{\min(\mathcal{C},\lambda \mathcal{C}_b)/\lambda}-1=2^{\min(\mathcal{C}/\lambda, \mathcal{C}_b)}-1$ as $\lambda$ is its time fraction. Thus, the observation below follows.
\begin{remark}
From Theorem \ref{TOA1}, it is seen that the \emph{ergodic rate} of the home user in open access is an increasing
function
of $\lambda_1$: As stated above $\Gamma_{f,1}$ dose not increase when $\lambda_1$ gets larger, then both
$F_I(\frac{P_f}{P_c \Gamma_{f,1}})$ and $\min(\mathcal{C},\lambda_1 \mathcal{C}_b)$ are non-decreasing functions w.r.t.
$\lambda_1$.
\end{remark}

The remark above implies that with a large enough value of $\lambda_1$, the home user's \emph{ergodic rate} can possibly
be higher than that in closed access. However, the following corollary shows that such a rate gain in open access is not
possible in the regime of very large cellular user density.
\begin{corollary}\label{coroTOA1}
\emph{If the values of $\lambda_1$ is independent of $N$, then as the number of cellular user goes to arbitrarily
large, that is $N \rightarrow \infty$,  the ergodic rate of home user in TDMA becomes}
\begin{align}
\mathcal{C}_0 =
\begin{cases}
\min(\mathcal{C}, \lambda_0 \mathcal{C}_b) F_I(\frac{P_f}{P_c \Gamma_{f,0}}) & \mbox{Closed Access} \\
\min(\mathcal{C},\lambda_1 \mathcal{C}_b) F_I(\frac{P_f}{P_c \Gamma_{f,1}})  & \mbox{Open Access}
\end{cases}
\end{align}
\end{corollary}

\begin{remark} \label{rehighN}
Since $\lambda_1 \leq \lambda_0=1$, $\Gamma_{f,1}$ is greater than $\Gamma_{f,0}$. Therefore, as shown in
Corollary \ref{coroTOA1}, with very high cellular user density, open access is inferior to closed access in terms of home
user's rate.
\end{remark}

When the number of cellular user increases, the time fraction of each loud neighbor decreases. Therefore, handing off a small group of loud neighbors (Corollary
\ref{coroTOA1} is derived in the case of $K=1$, but the argument can be extended) does not lower the macro-to-femto
interference significantly. Hence in the regime of high cellular user density, the loss of the FAP backhaul and time
resources becomes more dominant, leading to a decreased \emph{ergodic rate} of home user in open access. In other words,
the handoff metric in open access should not purely dependent on the distance between the interferers and the femtocells,
but also the time fraction they are active.

\section{Capacity Contours in Non-Orthogonal Multiple Access Scheme}
3G CDMA networks have been launched worldwide in recent years and will be in wide service for at least
a decade. This necessitates research and standardization for incorporating femtocells
in CDMA cellular networks \cite{DNK09CommMag, DNK09CommMag2,
PH09CommMag, MY09CommMag}. Even if both TDMA and CDMA are part of the medium access (e.g. HSPA in 3GPP and EVDO in 3GPP2), we restrict our attention to the CDMA aspect here, and this analysis would thus be valid per time or frequency slot. We show that in CDMA the interests of the femtocell owner and the network operator are compatible: Open access is the appropriate approach for both two parties.

In CDMA, suppose $L$ cellular users are served by the FAP, and the received SIRs at the two BSs are
    \begin{align} \label{equCSIR}
    \begin{cases}
    \frac{P_f}{L P_f + P_c \sum_{j=1}^{N-L} I_{(j)}}  & \mbox{SIR at the FAP} \\
    \frac{P_c}{P_f/I_0+ P_f \sum_{j=N-L+1}^N I_{(j)}+(N-L-1) P_c}  & \mbox{SIR at the macrocell BS}
    \end{cases}
    \end{align}
To be consistent with previous analysis in TDMA, we use the same notations of users' target SIRs, but note that their values change as the rate-SIR mapping function in CDMA is different due to spreading.
\begin{theorem}
\emph{In CDMA, the ergodic rate of the home user and the sum throughput of cellular users in femtocell closed access are
given
by}
\begin{flalign} \label{CCA}
\mathcal{C}_0=\min(\mathcal{C}, \lambda_0 \mathcal{C}_b) G_I(N,\frac{P_f}{P_c \Gamma_{f,0}}) \quad \quad
\mathcal{C}_{sum}=N \mathcal{C} \mathbb{P}\left(\frac{P_c}{P_f/I_0 + (N-1) P_c} \geq \Gamma_{c,0}\right)
\end{flalign}
\end{theorem}
\begin{IEEEproof}
In closed access, no cellular user is served by the FAP, meaning that the value of
$L$ in equation (\ref{equCSIR}) is zero. Thus, the success probabilities of the home user and the cellular users are
given by
\begin{flalign}
p_{f,0}=\mathbb{P} \left(\sum_{j=1}^N I_j \leq \frac{P_f}{P_c \Gamma_{f,0}}\right)=G_I\left(N,\frac{P_f}{P_c
\Gamma_{f,0}}\right) \quad \quad
p_{c,0}=\mathbb{P}\left(\frac{P_c}{P_f/I_0 + (N-1) P_c} \geq \Gamma_{c,0}\right)
\end{flalign}
Then the results of $\mathcal{C}_0$ and $\mathcal{C}_{sum}$ follow.
\end{IEEEproof}

Similar to TDMA or OFDMA, there exist cutoff user loadings $N_c^*$ and $N_o^*$ for \emph{sum throughput} in CDMA as well.
For example, in a Gaussian channel, the value $N_c^*$ is governed by
\begin{align}
\frac{2^{N_c^*\mathcal{C}}-1}{G}=\frac{P_c}{P_f/I_0+(N_c^*-1)P_c}
\end{align}

In femtocell open access, the mathematical expression of $\{A_L, L=0,1,2,\ldots,K\}$ is given by
\begin{align}
\mathbb{P} (A_L) =
\begin{cases}
\mathbb{P} \left(\bigcap\limits_{j=0}^{L-1}\left\{\frac{P_f}{j P_f + P_c \sum_{m=1}^{N-j}I_{(m)}} < \Gamma_{f,j}\right\},
\frac{P_f}{L P_f + P_c \sum_{m=1}^{N-L}I_{(m)}} \geq \Gamma_{f,L} \right) & L < K\\
\mathbb{P} \left( \bigcap\limits_{j=0}^{K-1} \left\{\frac{P_f}{j P_f + P_c \sum_{m=1}^{N-j}I_{(m)}} <
\Gamma_{f,j}\right\}\right) & L=K\\
\end{cases}
\end{align}
The general form of capacity contours in open access in CDMA are the same as those in Lemma \ref{TOA}, however in which
the success probabilities are different.
\begin{lemma} \label{COA}
\emph{In CDMA, the success probabilities for the home user, the supported cellular users at the femtocell and the remaining
cellular users at the macrocell are given by}
\begin{align}
p_{f,L}&=
\begin{cases}
\mathbb{P} (A_L) & L=0,1,\ldots,K-1 \\
\mathbb{P} \left(\frac{P_f}{P_c \sum_{m=1}^{N-K}I_{(m)} + K P_f} \geq \Gamma_{f,K}, A_K \right) & L=K\\
\end{cases} \end{align}
\begin{align}
p_{h,L}&=\mathbb{P} \left(\frac{P_f}{P_c \sum_{m=1}^{N-L}I_{(m)} + L P_f} \geq \Gamma_{h,L}, A_L \right) \\
p_{c,L}&=\mathbb{P} \left(\frac{P_c}{(N-L) P_c + P_f/I_0+ \sum_{m=N-L+1}^{N} P_f/I_{(m)}} \geq \Gamma_{c,L}, A_L \right)
\end{align}
\end{lemma}
\begin{IEEEproof}
The proof is very similar to Lemma \ref{TOA}, so is omitted.
\end{IEEEproof}
Based on Lemma \ref{COA}, we derive two helpful lower bounds in the following theorems.
\begin{theorem} \label{COA1}
\emph{In CDMA, the home user's ergodic rate in open access is
\begin{align}
\mathcal{C}_0 & = \min(\mathcal{C}, \lambda_0 \mathcal{C}_b) G_I(N,\frac{P_f}{P_c \Gamma_{f,0}})+ \sum_{L=1}^{K} \min
(\mathcal{C},\lambda_L \mathcal{C}_b) p_{f,L}
\end{align}
a lower bound on $p_{f,L}$ is given by
\begin{flalign}
p_{f,L} & \geq\left(1-L \binom{N}{L} B (x;
N-L+1,L)\right)G_I\left(N-L,\frac{P_f}{P_c}\left(\frac{1}{\Gamma_{f,L}}-L\right)\right)
\end{flalign}
}
\emph{Where $x=F_I(\frac{P_f}{P_c \Gamma_{f,L-1}})$, and the function $B(x;a,b)$ is the incomplete beta function}
\begin{align}
B(x;a,b)=\int_{0}^{x} t^{a-1} (1-t)^{b-1} dt
\end{align}
\end{theorem}
\begin{IEEEproof}
See Appendix \ref{COA1proof}.
\end{IEEEproof}
\begin{remark} \label{reCOA1}
Theorem \ref{COA1} shows that in CDMA, open access has strictly better performance than closed access in terms of the home
user's rate, irrespective of the femtocell resource allocation after handoff.
\end{remark}

This observation is quite different from that in TDMA. In CDMA, interference is additive. So handing off a small group of
strongest interferers always means a significant reduction in macro-to-femto interference and consequently an improvement in the home user's rate. On the contrary, interference is time shared in TDMA. Thus handing off a small group of interferers for just part of the time does not guarantee an appreciable reduction of cross-tier interference.

\begin{theorem} \label{COA2}
\emph{In CDMA, when the FAP is set to serve one cellular user at most, namely $K=1$, the sum
throughput of cellular users in open access is given by}
\begin{align}
\mathcal{C}_{sum} = N \mathcal{C} \mathbb{P}(\frac{P_c}{P_f/I_0 + (N-1)P_c} \geq \Gamma_{c,0})
G_I\left(N,\frac{P_f}{P_c \Gamma_{f,0}}\right) + (N-1)\mathcal{C} p_{c,1}+  \min(\mathcal{C}, \mu_1 \mathcal{C}_b)
p_{h,1}
\end{align}
\emph{where lower bounds of $p_{h,1}$ and $p_{c,1}$ are given by}
\begin{align}
p_{h,1} &\geq \left(1-F_I^N(\frac{P_f}{P_c \Gamma_{f,0}})\right)G_I \left(N-1,\frac{P_f (1-\Gamma_{h,1})}{P_c
\Gamma_{h,1}}\right)\\
p_{c,1} &\geq 1- F_I^N\left(y\right)
\end{align}
    \emph{where y is given by}
    \begin{align}
    y=\max\left(\frac{P_f \Gamma_{c,1}}{P_c-(N-2)P_c \Gamma_{c,1}- P_f \Gamma_{c,1}/I_0}, \frac{P_f}{P_c
    \Gamma_{f,0}}\right)
\end{align}
\end{theorem}
\begin{IEEEproof}
we first deploy the same technique as in the proof of theorem \ref{COA1} in deriving the lower bound of $p_{h,1}$
\begin{align}
p_{h,1}&=\mathbb{P} \left(\frac{P_f}{ P_c \sum_{m=1}^{N}I_{m}} < \Gamma_{f,0}, \frac{P_f}{P_f + P_c
\sum_{m=1}^{N-1}I_{(m)}} \geq \Gamma_{h,1} \right) \nonumber \\
& \geq \mathbb{P} \left(I_{(N)} >\frac{P_f}{P_c \Gamma_{f,0}} \right) G_I \left(N-1,\frac{P_f (1-\Gamma_{h,1})}{P_c
\Gamma_{h,1}}\right) \nonumber \\
&=\left(1-F_I^N(\frac{P_f}{P_c \Gamma_{f,0}})\right)G_I \left(N-1,\frac{P_f (1-\Gamma_{h,1})}{P_c \Gamma_{h,1}}\right)
\end{align}
A similar proof applies to the lower bound of $p_{c,1}$
\end{IEEEproof}

Similar to the statement in Remark \ref{resumT}, open access improves cellular users' \emph{sum throughput} in the regime of large $N$, while leading to a deterioration for small $N$. However, in CDMA the cause and the significance of such a \emph{sum throughput} loss are quite different. After the FAP serves $L$ cellular users, the femto-to-macro interference is $(N-L)P_c+P_f/I_0+\sum_{j=N-L+1}^N P_f/I_{(j)}$, smaller than $(N-1)P_c + P_f/I_0$ in closed access with high probability (according to Definition \ref{DefI} of ordered \emph{interference factors}). However, due to the resulting variance, the femto-to-macro interference in open access can exceed a certain threshold with a positive possibility, consequently causing outage to cellular users remaining in the macrocell. Open access thus causes a minor loss of \emph{sum throughput} in CDMA, as shown in Section V.

\begin{remark} \label{rsumC1}
In CDMA, open access is also the preferred choice for the network operator, since it is almost as good as (strictly better than) closed access in the regime of small (large) $N$.
\end{remark}
%

\section{Numerical Results $\&$ Conclusion}
Notations and system parameters are given in Table \ref{table1}. Note that in our plots,
the home user's \emph{ergodic rate} and cellular users' \emph{sum throughput} are normalized by $\mathcal{C}$.

\subsection{TDMA or OFDMA Access}
\begin{figure}[htp]
\centerline{
\includegraphics[width=3.5in]{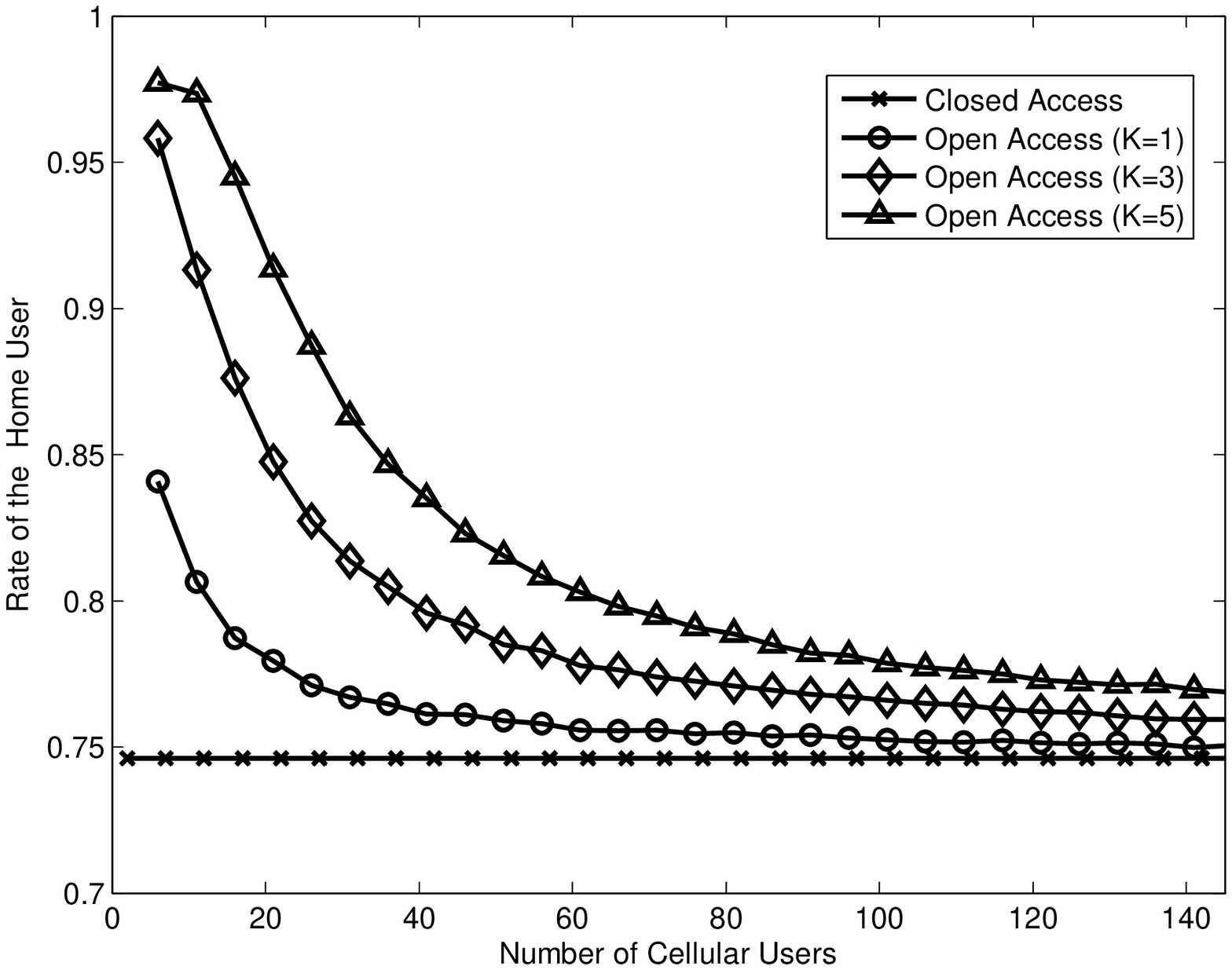}}
\caption{Home user's rate w.r.t. cellular user density in TDMA, $\lambda_L=1-\frac{L}{N}$ for fair comparison,
$\Gamma_{f,L}=2^{\min(0.5/\lambda_L,2)}-1$, $0 \leq L \leq K$.} \label{khomeTN}
\end{figure}

\begin{figure}[htp]
\centerline{\subfigure[Regime of small $N$]{\includegraphics[width=3in]{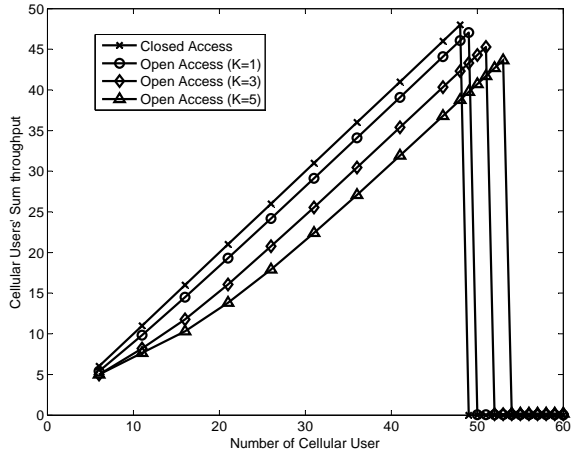} }
\hfil \subfigure[Regime of large $N$]{\includegraphics[width=3in]{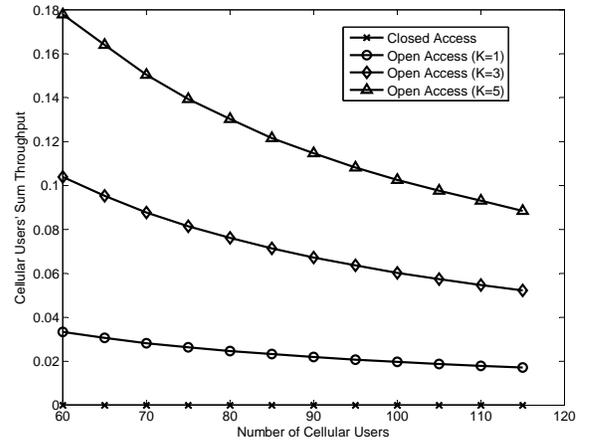} }}
\caption{Cellular users' sum throughput w.r.t. cellular user density in TDMA, $\mu_L=\frac{1}{N}$ for fair
comparison, $\Gamma_{h,L}=2^{\min(0.5 N,2)}-1$, $\Gamma_{c,L}=2^{0.5(N-L)}-1$, $0 \leq L \leq K$.} \label{ksumTN}
\end{figure}

\textbf{Cellular User Density.} Fig. \ref{khomeTN} and \ref{ksumTN} depict the home user's \emph{ergodic rate} and
cellular users' \emph{sum throughput} w.r.t. cellular user density. In low user density ($N \leq N_c^*=49$), open
access provides an appreciable rate gain to the home user, however which also causes a noticeable decrease in cellular users' \emph{sum throughput}. It is seen that the rate gain and loss are about the same in terms of percentage: For $K=3$ case at $N=20$, as an example, the rate gain of the home user is about 15\%, and the rate loss of cellular users is almost 20\%. Indeed, the choices of the two parties in low cellular user density are incompatible.

As predicted previously (below Remark \ref{reTCA}), there is a cutoff user loading $N_o^*$ in open access. Indeed, further simulations show as long as $\lambda \gtrsim 30\%$, $N_o^*$ equals $N_c^*+K$. In other words, single open access femtocell expands the macrocell network capacity by $K$, the largest amount of cellular users it can support. Therefore when $N_c^* \leq N \leq N_o^*$, open access is appropriate for both parties, especially to the network operator by offloading traffic from overloaded macrocell.

In high user density ($N \geq N_o^*=55$), open access provides very small rate gain to the home user, as predicted by
Corollary \ref{coroTOA1}. Additionally, it does not help network operator significantly because the macrocell BS is still
overloaded even after the femtocell serves $K$ cellular users for it. In other words, the \emph{sum throughput} in Fig
\ref{ksumTN} (b) is attributed to the $K$ cellular users served at the FAP. Considering the low rate gain and the potential security and privacy risks, there is not much motivation for open access in TDMA/OFDMA in high user density. We summarize these observations in Table
\ref{table2}.

\begin{figure}[ht]
\centerline{
\includegraphics[width=3.5in]{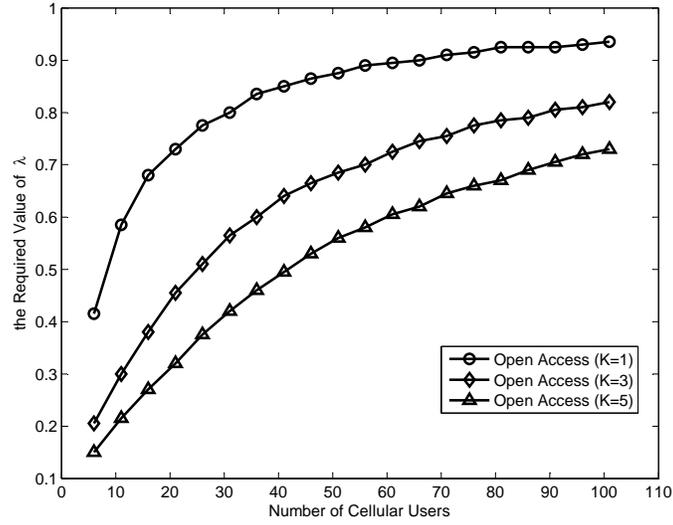}
}
\caption{The minimum femtocell resources required by home user in open access w.r.t. cellular user density in TDMA. The
home user's SIR target is $\Gamma_f=2^{\min(0.5/\lambda,2)}-1$}
\label{khomeTNlambda}
\end{figure}

\begin{figure} [ht]
\centerline{
\includegraphics[width=3.5in]{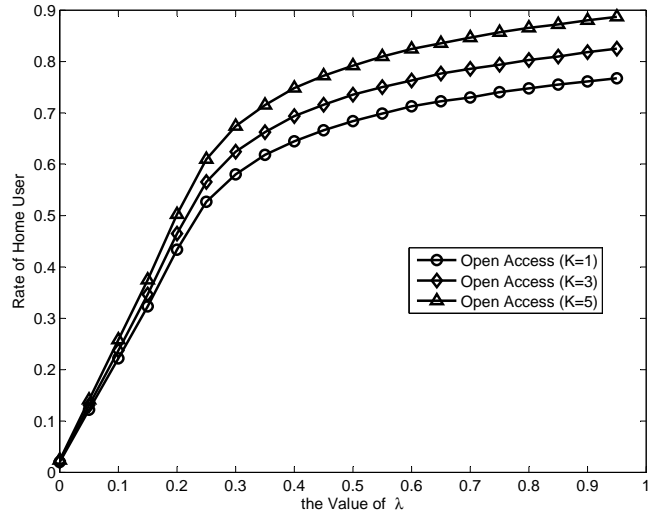}
}
\caption{Home user's rate w.r.t. the value of $\lambda$ in open access in TDMA. The home user's SIR target is
$\Gamma_f=2^{\min(0.5/\lambda,2)}-1$. $N=30$.} \label{kTlambda}
\end{figure}

\textbf{Femtocell Resource Allocation.} There exists a minimum value of $\lambda^*$ in femtocell resource allocation to ensure the home user benefits from open access\footnote{In our simulations, $\mu_L$ is set as $\frac{1-\lambda_{L}}{L}$. In this way, the limited femtocell backhaul is fully utilized. Additionally, analysis only on the value of
$\lambda_L$ will be sufficient to show the impact of femtocell resource allocation schemes.}. It is shown in Fig.
\ref{khomeTNlambda} that $\lambda^*$ is an increasing function of cellular user density $N$. Therefore the FAP resource allocation in TDMA or OFDMA should be adaptive to $N$, which is potentially difficult due to no coordination between the FAP and the macrocell BS. Otherwise, the performance of open access is very likely to degrade sharply, because the home user's \emph{ergodic rate} is sensitive to the value of $\lambda$, as shown in Fig. \ref{kTlambda}.

\begin{figure}[ht]
\centerline{
\includegraphics[width=3.5in]{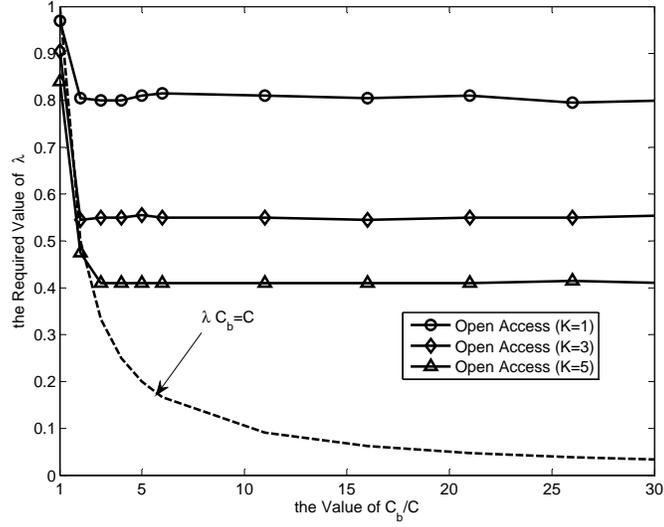}
}
\caption{The minimum femtocell resources required by the home user in open access w.r.t. femtocell backhaul constraint in
TDMA. The home user's SIR target is $\Gamma_f=2^{\min(0.5/\lambda,2)}-1$. $N=30$.} \label{khomeTNCblambda}
\end{figure}

The explanation to above observations are given as follows. Although the performance of open access is governed by both
backhaul and time allocation in the FAP, the latter factor is more dominant: As long as $\mathcal{C}_b$ is not very small
(i.e. $\mathcal{C}_b/\mathcal{C} \geq 3$ in Fig \ref{khomeTNCblambda}), the value of $\lambda^*$ (the solid curves) are constants and high above the dashed line, which is governed by $\lambda \mathcal{C}_b = \mathcal{C}$, denoting the backhaul limitations. This implies the value of $\lambda^*$ is mainly determined by home user's need of time resources, instead of backhaul capacity. This explains why in TDMA or OFDMA $\lambda^*$ is dependent on $N$ and open access is very sensitive to FAP resource allocation.

\textbf{Summary for TDMA/OFDMA Access.} In orthogonal multiple access, the choices of the two parties are highly dependent on the cellular user density, with both preferring open access in medium density, closed access in high density, and they are in disagreement at low density. Therefore, our results suggest that when deploying OFDMA, femtocell access control should be adaptive based on the estimated cellular user density.  Note that this conclusion is probably contingent on the assumption of no coordination between the femtocells and the macrocell BS.  Future work should (and surely will) consider inter-BS coordination both among and across the two tiers, which is surely helpful and our conjecture will be especially important to open access in high user densities.

\subsection{CDMA Access}
\begin{figure}[ht]
\centerline{
\includegraphics[width=3.5in]{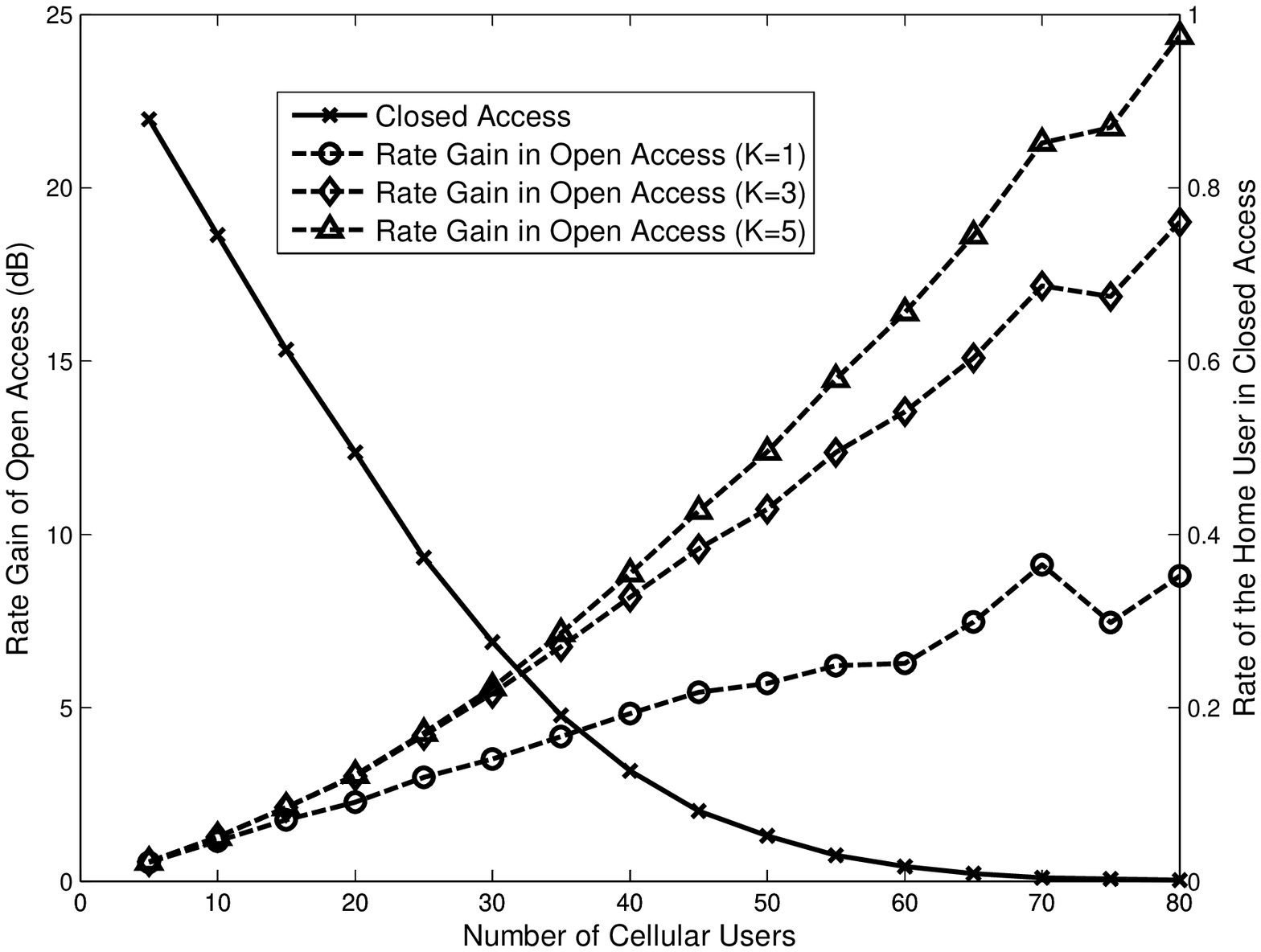}
}
\caption{Home user's rate gain w.r.t. cellular user density in CDMA, $\lambda_L=1-\frac{L}{N}$ for fair comparison,
$\Gamma_{f,L}=(2^{\min(0.5,2\lambda_L)}-1)/G$, $0 \leq L \leq K$.} \label{khomeCN}
\end{figure}

\begin{figure}[ht]
\centerline{
\includegraphics[width=3.5in]{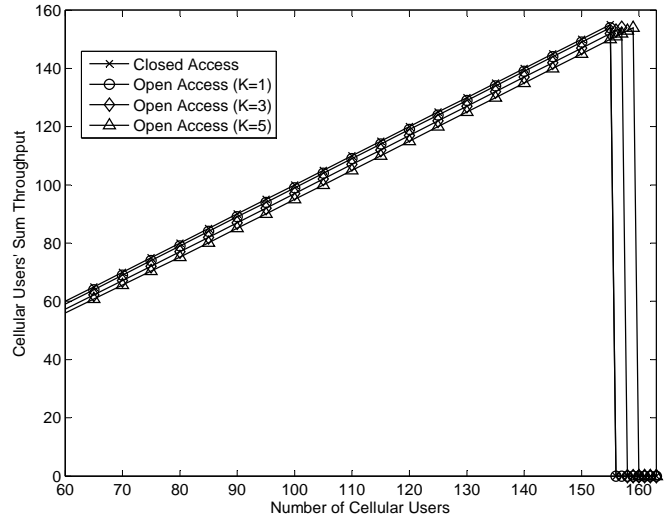}
}
\caption{Cellular users' sum throughput in low cellular user density in CDMA, $\mu_L=\frac{1}{N}$ for fair
comparison, $\Gamma_{c,L}=(2^{\min(0.5,2\mu_L)}-1)/G$, $0 \leq L \leq K$.} \label{ksumCN}
\end{figure}

\textbf{Cellular User Density.} Theorem \ref{COA1} states that the home user will always experience a rate gain in open
access in CDMA, which is over 200\% (5 dB) for a vast range of cellular user density, as shown in Fig.
\ref{khomeCN}. As predicted by Remark \ref{rsumC1}, Fig. \ref{ksumCN} shows that in the regime of small $N$ ($N \leq
N_c^* =155$), open access in CDMA only leads to a negligible loss of cellular users' \emph{sum throughput}. Open access in CDMA will strictly improve cellular users' \emph{sum throughput} when $N \geq N_c^*=155$, for the same reason in TDMA or
OFDMA. Therefore, in CDMA open access is an appropriate approach for both parties in the whole range of cellular user density.

\textbf{Femtocell Resource Allocation.} As stated in Theorem \ref{COA1} and Remark \ref{reCOA1}, open access improves the
home user's rate, no matter how the femtocell backhaul is shared among users and what the cellular user density is.
Indeed, as long as $\lambda \mathcal{C}_b \geq \mathcal{C}$, the home user's rate is not affected by the femtocell
resource allocation, and the rate gain in Fig. \ref{khomeCN} will be achieved.

\textbf{Summary for CDMA Access.} Open access in CDMA benefits both parties in almost the whole range of cellular user density. Moreover, these appreciable benefits do not require the FAP to deploy adaptive resource allocation. Therefore, open access is conclusively preferred in 3G CDMA networks.

%
\appendix
\subsection{Proof of Lemma \ref{cdf}} \label{cdfproof}
Denote $(x,y)$ as the location of a cellular user. Thus the CDF of its \emph{interference factor} $I$ is
\begin{align}
F_I(i) &= \mathbb{P}(I \leq i) \nonumber \\
&= \mathbb{P} \left((1-i^{2/\alpha})x^2+(1-i^{2/\alpha})y^2+2D i^{2/\alpha}x-D^2 i^{2/\alpha}
\leq 0 \right) \nonumber \\
&= S/(\pi R^2).
\end{align}
$S$ is the area inside the macrocell, and governed by
\begin{align}
(1-i^{2/\alpha})x^2+(1-i^{2/\alpha})y^2+2D i^{2/\alpha}x-D^2 i^{2/\alpha} \leq 0 \nonumber
\end{align}
When $i \neq 1$, the above equation defines a circle area, with center of $x_c=\frac{D i^{2/\alpha}}{|1-i^{2/\alpha}|}$
and radius of $r=\frac{i^{1/\alpha}D}{|1-i^{2/\alpha}|}$. Moreover, when $i <1$, $S$ is the area inside the circle, while
$i >1$, $S$ is the area outside the circle. Therefore, according to the range of $i$, $F_I(i)$
can be divide into 5 segments:
\begin{enumerate}
\item when $0 \leq i <(\frac{R}{R+D})^{\alpha}$  (note that \emph{interference factor} is a non-negative r.v., we
    must
    have $i \geq 0$), the circle is contained in macrocell ($|x_c|+|r| \leq R$). Thus $S=\pi r^2$.
\item when $(\frac{R}{R+D})^{\alpha} \leq i < 1$, the circle intersects with macrocell ($|x_c|+|r| \geq R$). Using
    the
    method in \cite{JZ08Twc}, we get:
\begin{equation}
S=\pi(R^2+r^2)-\left(\pi-\theta+0.5 \sin 2\theta\right)r^2-(\pi-\phi+0.5\sin2\phi) R^2 \nonumber
\end{equation}
where $\theta=\arccos(\frac{r^2+x_c^2-R^2}{2 r x_c})$ and
$\phi=\arccos(\frac{x_c^2+R^2-r^2}{2 x_c R})$.
\item when $i=1$, the area of $S$ is a half plane ($x \leq \frac{D}{2}$) intersected with the macrocell. Thus $S=(\pi
    -
    \varphi + 0.5
\sin2\varphi) R^2$, where $\varphi=\arccos(\frac{D}{2 R})$.
\item when $1 < i \leq (\frac{R}{R-D})^{\alpha}$, the circle intersects with macrocell ($|x_c|+|r| \geq R$). Note
    that
    now $S$ is the area outside the circle. Therefore
\begin{equation}
S=\pi R^2-\pi(R^2+r^2)+\left(\pi-\theta+0.5 \sin2\theta\right)r^2+(\pi-\phi+0.5\sin2\phi) R^2 \nonumber
\end{equation}
\item when $(\frac{R}{R-D})^{\alpha} < i$, the circle is contained in macrocell ($|x_c|+|r| \leq R$). Similarly,
    $S=\pi (R^2-r^2)$.
\end{enumerate}

\subsection{Proof of Theorem \ref{TOA1}} \label{TOA1proof}
As $K=1$, there are only two events $A_0$ and $A_1$, with probabilities of
\begin{align}
\mathbb{P} (A_0)=F_I^N(\frac{P_f}{P_c \Gamma_{f,0}}) \quad \mathbb{P} (A_1)=1-F_I^N(\frac{P_f}{P_c \Gamma_{f,0}})
\end{align}
The key in the proof is the calculation of $p_{f,1}$,
$p_{h,1}$ and $p_{c,1}$. Applying (\ref{equTPi}), we have
\begin{align} \label{equ1}
p_{f,1}&= \frac{1}{N-1} \sum_{j=1}^{N-1} \mathbb{P}\left(I_{(j)} \leq \frac{P_f}{P_c \Gamma_{f,1}},I_{(N)} >
\frac{P_f}{P_c \Gamma_{f,0}}\right)
\end{align}
Use the bins and balls technique, we denote $\mathbb{P}(I \leq \frac{P_f}{P_c \Gamma_{f,1}})$ as $p$, and $\mathbb{P}(I
\leq \frac{P_f}{P_c \Gamma_{f,0}})$ as $q$ to solve
\begin{align}
& \sum_{j=1}^{N-1} \mathbb{P}(I_{(j)} \leq \frac{P_f}{P_c \Gamma_{f,1}},I_{(N)} > \frac{P_f}{P_c \Gamma_{f,0}}) \nonumber
\\
=& \sum_{j=1}^{N-1} \sum_{k=j}^{N-1} \left\{\binom{N}{k} p^k \left(1-p\right)^{N-k}- \binom{N}{k} p^k
\left(q-p\right)^{N-k} \right\} \nonumber \\
=& N p (1-q^{N-1})
\end{align}
Substituting back for $p$ and $q$, $p_{f,1}$ is given by
\begin{align}
    p_{f,1}&=\frac{N}{N-1} F_I(\frac{P_f}{P_c \Gamma_{f,1}}) \left( 1- F_I^{N-1}(\frac{P_f}{P_c \Gamma_{f,0}})\right)
\end{align}
The success probability $p_{h,1}$ of handoff user $U_{(N)}$ follows by applying the same technique.

In the femtocell, the home user is allocated a time fraction of $\lambda_1$, and the handed off user $U_{(N)}$ is
assigned a time fraction of $\mu_1$. Therefore, a cellular user in the macrocell network will experience interference
from the home user with probability $\lambda_1$ and from the handed off user $U_{(N)}$ with probability $\mu_1$. Therefore, its success probability is
\begin{align}
p_{c,1}&= \lambda_1 \mathbb{P} \left(\frac{P_c}{P_f/I_0} \geq \Gamma_{c,1}\right) + \mu_1 \mathbb{P}
\left(\frac{P_c}{P_f/I_{(N)}} \geq \Gamma_{c,1}, \frac{P_f}{P_c I_{(N)}} < \Gamma_{f,0}\right) \nonumber \\
&=\lambda_1 \mathbb{P}\left(\frac{P_c}{P_f/I_0} \geq \Gamma_{c,1}\right)+ \mu_1 \min\left\{1-F_I^N(\frac{P_f}{P_c
\Gamma_{f,0}}),1-F_I^N(\frac{P_f \Gamma_{c,1}}{P_c})\right\}
\end{align}

\subsection{The Proof of Theorem \ref{COA1}} \label{COA1proof}
As $L=0$, we have
\begin{align}
p_{f,0}=\mathbb{P} \left(\frac{P_f}{P_c \sum_{m=1}^{N}I_{(m)}} \geq \Gamma_{f,0}\right)=G_I \left(N, \frac{P_f}{P_c
\Gamma_{f,0}}\right)
\end{align}
For $1 \leq L \leq K$, it is easy to check that $p_{f,L}$ has the same form, which are lower bounded as below.
\begin{align}
p_{f,L}&=\mathbb{P} \left(\bigcap\limits_{j=0}^{L-1} \left\{\frac{P_f}{j P_f + P_c \sum_{m=1}^{N-j}I_{(m)}} <
\Gamma_{f,j}\right\}, \frac{P_f}{L P_f + P_c \sum_{m=1}^{N-L}I_{(m)}} \geq \Gamma_{f,L} \right) \nonumber \\
&\geq \mathbb{P} \left(\bigcap\limits_{j=0}^{L-1} \left\{\sum_{m=1}^{N-j}I_{(m)} >\frac{P_f}{P_c \Gamma_{f,j}}\right\},
\sum_{m=1}^{N-L}I_{(m)} \leq \frac{P_f (1-L\Gamma_{f,L})}{P_c \Gamma_{f,L}} \right) \nonumber \\
& \stackrel{(a)}{\geq} \label{a} \mathbb{P} \left(\sum_{m=1}^{N-L+1}I_{(m)} >\frac{P_f}{P_c \Gamma_{f,L-1}},
\sum_{m=1}^{N-L}I_{(m)} \leq \frac{P_f (1-L\Gamma_{f,L})}{P_c \Gamma_{f,L}} \right) \nonumber \\
& \geq \mathbb{P} \left(I_{(N-L+1)} >\frac{P_f}{P_c \Gamma_{f,L-1}}, \sum_{m=1}^{N-L}I_{(m)} \leq \frac{P_f
(1-L\Gamma_{f,L})}{P_c \Gamma_{f,L}} \right) \nonumber \\
& \stackrel{(b)}{\geq} \mathbb{P} \left(I_{(N-L+1)} >\frac{P_f}{P_c \Gamma_{f,L-1}} \right) G_I \left(N-L,\frac{P_f
(1-L\Gamma_{f,L})}{P_c \Gamma_{f,L}}\right) \nonumber \\
& = \left[1-L \binom{N}{L} B \left(F_I(\frac{P_f}{P_c \Gamma_{f,L-1}});
N-L+1,L\right)\right]G_I\left(N-L,\frac{P_f}{P_c}\left(\frac{1}{\Gamma_{f,L}}-L\right)\right)
\end{align}
It is important to note that in CDMA $\Gamma_{f,L}$ is a non-decreasing function of $\lambda_L$. In a Gaussian channel, for
example, $\Gamma_{f,L}=\left(2^{\min(\mathcal{C},\lambda_L \mathcal{C}_{sum})}-1\right)/G$. Since $\lambda_L$ does not
increase as $L$ goes larger, $\Gamma_{f,L}$ is also a non-increasing function of $L$. Thus the inequality (a) holds.
Instead of making $\sum_{m=1}^{N-L}I_{(m)}$ smaller than a certain constant, we randomly pick $N-L$ elements from the set
of $\{I_{(1)},I_{(2)},\ldots,I_{(N)}\}$, which are also independent of $I_{(N-L+1)}$. The sum of these randomly picked
elements is larger than $\sum_{m=1}^{N-L}I_{(m)}$. Therefore inequality (b) holds.

\bibliographystyle{IEEEtran}
\bibliography{FEMTO}

\newpage

\begin{table}[ht]
\caption{notations and parameters} \centering
\begin{tabular}{c|c|c}
\hline \hline Symbol & Description & Sim. Value\\
\hline
\hline $R$ & macrocell radius & $300$ meters\\
\hline $D$ & Distance between macro \& femto base stations & $150$ meters\\
\hline $d$ & Distance between home user and femtocell BS & $5$ meters\\
\hline $\alpha, \beta$ & Path loss exponents & $4,2$\\
\hline $I$ & Ratio of user's channel to the FAP over that to macrocell BS & N/A\\
\hline $P_f, P_c$ & Femtocell \& macrocell BS receive power & $1,1$\\
\hline $G$ & Spreading factor (for CDMA)& $64$\\
\hline $\mathcal{C}, \mathcal{C}_b$ & User rate requirement $\&$ Femtocell backhaul capacity& $0.5$ bps/Hz $\&$ $2$
bps/Hz\\
\hline $\mu_L$ & Portion of FAP resources allocated to each handed off cellular user & $1/N$ (for fair comparison)\\
\hline $\lambda_L$ & Portion of FAP resources allocated to the home user & $1-\frac{L}{N}$ with $L$ handed off cellular
users\\
\hline \multirow{2}*{$\Gamma_{f,L}$} & \multirow{2}*{SIR target of the home user} &
$2^{\min(\mathcal{C}/\lambda_L,\mathcal{C}_b)}-1$ (TDMA)\\ & & $(2^{\min(\mathcal{C},\lambda_L \mathcal{C}_b)}-1)/G$
(CDMA)\\
\hline \multirow{2}*{$\Gamma_{h,L}$} & \multirow{2}*{SIR target of each handed off cellular user} &
$2^{\min(\mathcal{C}/\mu_L,\mathcal{C}_b)}-1$ (TDMA)\\ && $(2^{\min(\mathcal{C},\mu_L \mathcal{C}_b)}-1)/G$ (CDMA)\\
\hline \multirow{2}*{$\Gamma_{c,L}$} & \multirow{2}*{SIR target of each cellular user in the macrocell} &
$2^{(N-L)\mathcal{C}}-1$ (TDMA)\\ & & $(2^\mathcal{C}-1)/G$ (CDMA)\\
\hline $\mathcal{C}_0$ & \emph{Ergodic rate} of the home user & N/A \\
\hline $\mathcal{C}_{sum}$ & \emph{Sum throughput} of cellular users & N/A \\
\hline $N_c^*, N_o^*$ & Cutoff user loadings for \emph{sum throughput} in closed \& open access & N/A \\
\hline\hline
\end{tabular} \label{table1}
\end{table}

\begin{table} [htp]
\caption{choices of two parties w.r.t. cellular user density
}
\centering
\begin{tabular} {c|c|c|c|c}
\hline \hline & \multicolumn{2}{c|}{TDMA or OFDMA} & \multicolumn{2}{c}{CDMA}\\
\hline & Femtocell Owner & Network Operator & Femtocell Owner & Network Operator \\
\hline
\hline Low Cellular User Density ($N < N_c^*$)& Open Access & Closed Access & Open Access & Indifferent\\
\hline Medium Cellular User Density ($N_c^* \leq N \leq N_o^*$)& Open Access & Open Access & Open Access & Open Access\\
\hline High Cellular User Density ($N > N_o^*$)& Closed Access & Indifferent & Open Access & Indifferent\\
\hline Choices & \multicolumn{2}{c|}{Highly Dependent on Cellular User Density} & \multicolumn{2}{c}{Open Access}\\
\hline \hline
\end{tabular} \label{table2}
\end{table}

\end{document}